# Growth of Self Organized Eutectic Fibers from LiF─Rare Earth Fluoride Systems

Detlef Klimm[1], Maria F Acosta[2,1], Ivanildo A dos Santos[3,1], Izilda M Ranieri[3], Steffen Ganschow[1], and Rosa I Merino[2]

[1]Leibniz Institute for Crystal Growth, Max-Born-Str. 2
12489 Berlin, Germany

[2]Instituto de Ciencia de Materiales de Aragón (ICMA), Universidad de Zaragoza - CSIC,
50009 Zaragoza, Spain.

[3]Instituto de Pesquisas Energéticas e Nucleares, CP 11049, Butantã
05422-970 São Paulo, SP, Brazil

## ABSTRACT

Eutectic fibers consisting of an ordered arrangement of LiF fibrils inside a $LiREF_4$ matrix (RE = Y, Gd) can be grown with the micro-pulling-down method at sufficiently large pulling rate exceeding 120 mm/h. The distance between individual fibrils could be scaled down to 1 µm at 300 mm/h pulling. $LiF$-$LiYF_4$ has stronger tendency to form facetted eutectic colonies than $LiF$-$LiGdF_4$, explained by the larger entropy of melting of the former.

## INTRODUCTION

In eutectic systems $\{x \text{ A} + (1-x) \text{ B}\}$ at the eutectic composition $x_{eut}$ ($0 \leq x_{eut} \leq 1$) both components A and B are crystallizing simultaneously at the eutectic temperature $T_{eut}$. As the eutectic is an invariant point of the corresponding system, the shares of both components in the eutectic microstructure are almost fixed. Just minor variations from the eutectic composition $x_{eut}$ of a few per cent are typically allowed – if the deviation is larger, first the pure excess component crystallizes until the melt composition approaches the eutectic.

The morphology of eutectic microstructures depends on a variety of parameters, such as the volume fraction $xV_A / \{ xV_A + (1-x)V_B \}$ of the constituents ($V_A$, $V_B$ are the molar volumes of the components), the entropies of fusion, the thermal gradients $G_T$ at the solid-liquid interface, and on the solidification rate $v$. Only $v$ and $G_T$ are experimental parameters that can be chosen within certain limits almost arbitrarily for a given system, whereas the other parameters depend only on the substances themselves. If λ is the average distance between neighboring particles of one component, then one can show that the product $\lambda^2 v$ is constant for a given system. For sufficiently large $v$ one can expect motif scaling down to microns, making such eutectics interesting e.g. for photonic applications. In the THz range alkali halide ordered fibrous eutectics have been studied as polaritonic metamaterials [1], showing hyperbolic dispersion relations and thus potential for sub-wavelength resolution and THz imaging. The suitable wavelength depends on the materials combination chosen so that exploration of other eutectics that achieve ordered microstructures and hyperbolic dispersion at different wavelengths is of interest. At optical wavelengths (VIS and NIR), light guiding is expected [2]. For a more comprehensive introduction to the directional solidification of eutectics in general, the reader is referred to Orera et al. [3].

Barta et al. [4] performed a quantitative analysis of the microstructure of LiF−LiYF$_4$ eutectics that were directionally crystallized in graphite crucibles with conical tip and 15 mm diameter (Bridgman method). This large sample diameter, together with the substantial transport of heat through the well conducting crucible wall, restricted thermal gradients $G_T$ to 30–80 K/cm. Growth rates between 6 and 20 mm/h were used in that work.

In this study micro-pulling-down (µ-PD) is used for LiF−LiYF$_4$ and LiF−LiGdF$_4$ eutectics as an alternative method. The eutectic rods produced by this technique are much thinner; typically the diameter is well below 2 mm. The significant scale reduction enabled larger temperature gradients up to several 100 K/cm and consequently growth rates up to several 100 mm/h.

The change from Bridgman to µ-PD is not straightforward, because the surface/volume ratio of thin rods is much larger, and the surface is more exposed. This can result in severe contamination because fluorides are sensitive against hydrolysis.

## EXPERIMENT
### Phase diagrams

The phase diagrams LiF−YF$_3$ and LiF−GdF$_3$ were described first by Thoma et al. [5,6] and later basically confirmed by the group around Sobolev [7,8]. The eutectics that are studied here are situated between the intermediate scheelite type LiREF$_4$ (RE = Y, Gd) and LiF. Both systems LiF−YF$_3$ [9] and LiF−GdF$_3$ [10] were re-investigated and for the first time a thermodynamic assessment for the Gibbs free energies $G(T)$ of the line compounds LiF, YF$_3$, LiYF$_4$ and $G(T,x)$ of the $\{x\, YF_3 + (1-x)\, LiF\}$ melt was performed. The scheelite type intermediate phase is a "borderline peritectic" where the liquidus of both YF$_3$ and LiYF$_4$, as well as the peritectic line of LiYF$_4$ meet at one point. LiGdF$_4$ forms a real peritectic: the LiF−GdF$_3$ has a topology similar like Fig. 1, with the difference that the GdF$_3$ liquidus and the LiGdF$_4$ peritectic line (at 1028 K) extend to $x = 0.34$, beyond the LiGdF$_4$ composition.

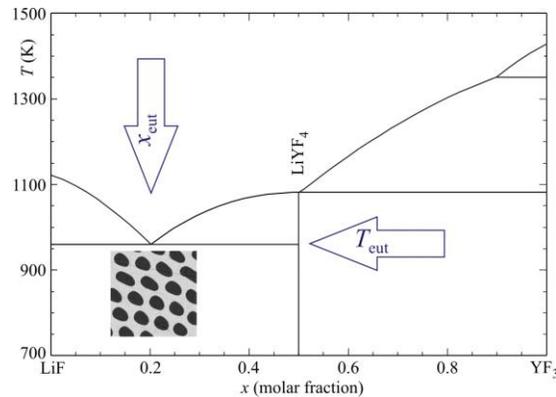

Fig. 1: Assessed phase diagram LiF-YF$_3$ with $T_{eut}$ = 975 K (702°C) and $x_{eut}$ = 0.202 [9]. The inset shows self-ordered LiF fibrils inside a LiYF$_4$ matrix forming during solidification at composition $x_{eut}$.

## Fiber growth

Eutectic LiF/LiYF$_4$ and LiF/LiGdF$_4$ rods were grown in a μ-PD apparatus that was recently used for the growth of LiYF$_4$ single crystals [11]. The whole setup (Fig. 2) is placed inside a 35 liter vacuum-tight steel chamber that is evacuated prior to growth below $10^{-5}$ mbar. It should be noted that rare earth fluorides are highly sensitive against hydrolysis with traces of moisture. Extreme dry conditions during all heat treatments, such as growth from the melt, are mandatory. Ar atmosphere (1 bar, 99.999% purity, <1 ppm water) was used.

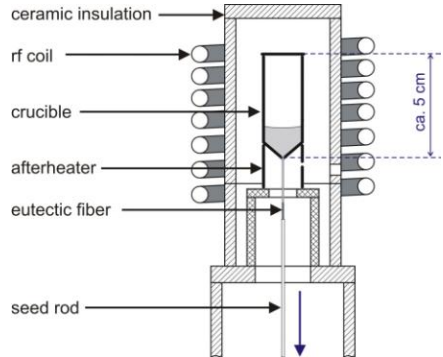

Fig. 2: Micro-Pulling-Down (μ-PD) growth of eutectic fibers.

Lithium fluoride, yttrium fluoride and gadolinium fluoride with 99.99% purity were mixed in appropriate ratio to form the corresponding eutectic mixture. The purity, and especially the absence of significant oxygen contamination, was controlled by DTA measurements and the phase transformation or melting temperatures found there were in good agreement with literature data, proving good purity (Fig. 3). A platinum wire was used as seed. Pulling rates ranging from 15 to 300 mm/h were used for the μ-PD experiments, which extends significantly the range that is accessible by the Bridgman method (4–60 mm/h).

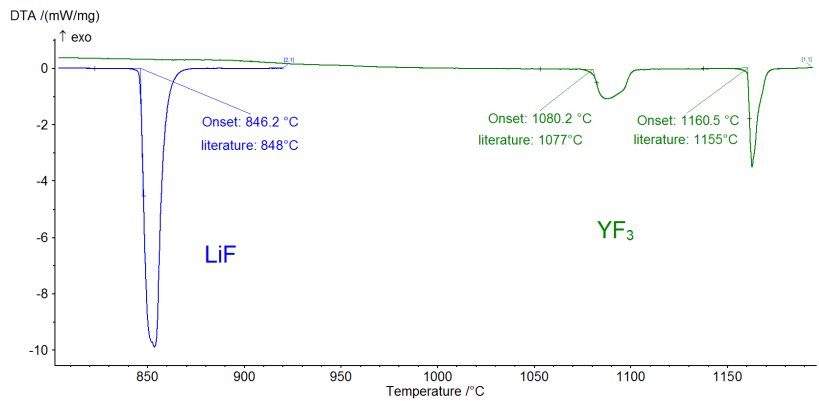

Fig. 3: DTA heating curves of the starting materials LiF (melting) and YF$_3$ (phase transformation and melting). (Measured with a NETZSCH STA 449C with vacuum-tight Pt/Rh furnace, literature data from FactSage [12]).

## RESULTS AND DISCUSSION

Fig. 4 shows SEM micrographs of transverse cross-sections of LiF-LiGdF$_4$ samples grown by the micro-pulling down method. For slow pulling rates between 15 and 60 mm/h, a coupled interpenetrated microstructure was found for both systems, LiF-LiYF$_4$ and LiF-LiGdF$_4$. SEM images reveal a transition in the microstructure when increasing the pulling rate from coupled interpenetrated to macrofacetted (Fig. 4a and Fig. 4c) in both systems. These macrofacetted cells consist of LiF rods embedded in a LiYF$_4$ or LiGdF$_4$ matrix, respectively. The crossover pulling rate occurs between 60 and 120 mm/h for the LiF-LiYF$_4$ system whereas for the LiF-LiGdF$_4$ system it is found at faster pulling rates between 120 and 300 mm/h. Areas of fibrilar ordered arrangements at least 100×200 µm$^2$ large have been observed. In the LiF-LiGdF$_4$ eutectic there is a range of pulling rates (between 120 and 300 mm/h) where LiF rods inside LiGdF$_4$ matrix are observed.

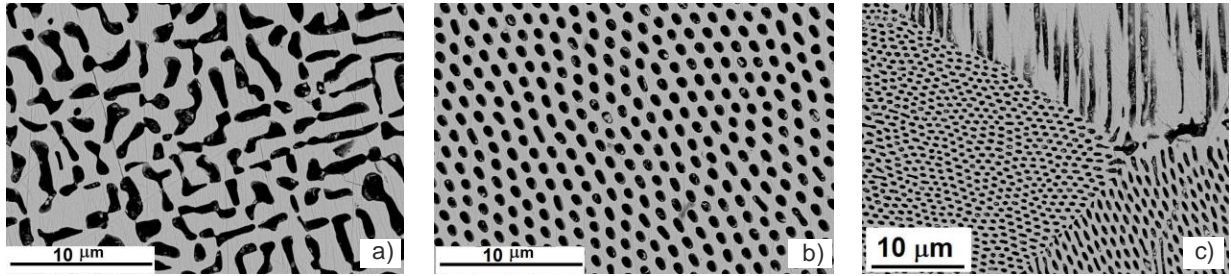

Fig. 4: SEM images of transverse cross-sections of LiF-LiGdF$_4$ rods grown at 60 mm/h (a), 120 mm/h (b) and 300 mm/h (c) by the µ-PD method. The coupled interpenetrated microstructure (a) changes into a macrofacetted one with a fibrilar arrangement (c) when increasing the pulling rate. Dark phases correspond to LiF.

Phase interspacing was obtained from SEM images by using the software Digital Micrograph from Gatan Inc.) The experimental data can be fitted by the empirical Jackson-Hunt law $\lambda^2 v = K$ (Fig. 5). Values obtained for $K$ were 106.1±0.2 µm$^3$/s for the LiF-LiYF$_4$ and 87.15±0.04 µm$^3$/s for the LiF-LiGdF$_4$ system. The nominal volumetric fractions were 40vol% LiF for the LiF-LiYF$_4$ and 33vol% LiF for the LiF-LiGdF$_4$ system, in good accord with the estimates from image analysis. The interfiber spacing could be scaled from 5 down to 1 µm by increasing the pulling rate from 15 mm/h to the maximum accessible value 300 mm/h. Short-range ordering between the LiF fibers is usually good even for the largest pulling rates, with change in orientation of the triangular lattice of rods that is commonly found in rod eutectics.

The $K$ values of the Jackson-Hunt relationship are very similar for both eutectics. The difference in the slopes of the liquidus curves around the eutectic point might be enough to substantiate the 20% larger $K$ for LiF-LiYF$_4$, even without taking into account possible differences in diffusion constant or surface tension between the components.

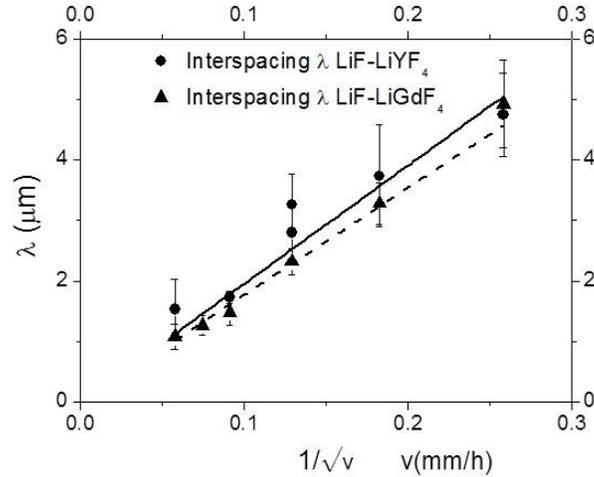

Fig. 5: Interfiber spacing as a function of the square root of the pulling rate for the eutectic systems LiF-LiYF$_4$ (circles) and LiF-LiGdF$_4$ (triangles) grown by the µ-PD method.

Even if both eutectics are very similar (concerning microstructure and interphase spacing), the formation of macrofacetted cells (characteristic of uncoupled growth) sets on earlier (at slower pulling rates) in LiF-LiYF$_4$ than in LiF-LiGdF$_4$. Usually, the question whether non-facetted or facetted growth appears is discussed in term of the Jackson parameter α ≈ $\Delta S/R$ ($\Delta S$ – entropy of fusion, $R$ = gas constant). For α ≤ 2 almost isotropic (non-facetted) growth can be expected [3]. The transformation of these ideas to eutectics with more than one component is not straightforward; especially if one of the components melts under peritectic decomposition in its pure form (LiGdF$_4$), and $\Delta S$ is not well defined. Fortunately at the eutectic temperature the eutectic mixture is in direct equilibrium with the melt, without formation of the rare earth fluoride. Thus, one can define a $\Delta S'$ for the eutectic itself by calculating the difference of $S$ for the mixture $x_{eut}$ just above and below $T_{eut}$. For LiF/LiGdF$_4$ this was performed with the assessment data for the LiF-GdF$_3$-LuF$_3$ system [13] and data for the LiF-YF$_3$ system where published recently [9]. For LiF-LiGdF$_4$ one calculates $\Delta S'$ = 28.2 J/(mol·K) = 3.4$R$ and for LiF-LiYF$_4$ $\Delta S'$ = 34.4 J/(mol·K) = 4.1$R$. The larger value of $\Delta S'$ for the LiF-LiYF$_4$ can explain why the faceting is more pronounced in this system. It should be noted that the values for both systems are large, compared e.g. with intermetallics: For Pb-Sn one finds at the eutectic point (36.1% Sn, 151°C) $\Delta S'$ = 15.1 J/(mol·K) = 1.8$R$ (FactSage data [12]).

Further research of the crystallographic orientation relationships and their influence on the microstructure will be done. The objective is to find out the growth conditions to achieve fibers with large domains of aligned rod-like or interpenetrated microstructures with sizes that allow tailoring light propagation in the material. A preliminary evaluation of the THz behavior of LiF-LiYF$_4$ composite has been done [14]. For this composition also selective etching of the LiF phase has been observed.

**CONCLUSIONS**

The growth of eutectic fibers from lithium fluoride – rare earth fluoride systems is possible, if very dry growth conditions are used, avoiding hydrolysis. With sufficiently large pulling rates >120 mm/h self-organized ordering of LiF fibrils inside a LiREF$_4$ matrix (RE = Y,

Gd) occurs. The interphase spacing follows the Jackson-Hunt rule and can be scaled down to ca. 1 μm with 300 mm/h pulling rate; options for further minimization with even faster pulling cannot be ruled out. The stronger tendency to faceting of LiF-LiYF$_4$ can be explained by the larger entropy of melting of this eutectic.

## ACKNOWLEDGMENTS

This work received financial support from the European Union (7[th] Framework Programme, "ENSEMBLE", NMP4-SL-2008-213669), from CNPq (477595/2008-1; 290111/2010-2) and from DAAD-CAPES (po-50752632). MFA acknowledges Ministerio de Educación, Cultura y Deporte (Spain) for the FPU scholarship.